\documentclass{amsproc}

\usepackage{graphicx}

\theoremstyle{definition}

\theoremstyle{remark}

\numberwithin{equation}{section}

\newcommand{\bq}{\begin{eqnarray}}
\newcommand{\eq}{\end{eqnarray}}
\newcommand{\eps}{\varepsilon}

\begin{document}

\title{Periods and Hodge structures in perturbative quantum field theory}

\author{Stefan Weinzierl}
\address{PRISMA Cluster of Excellence, Institut f{\"u}r Physik, Johannes Gutenberg-Universit{\"a}t Mainz, D - 55099 Mainz, Germany}
\email{stefanw@thep.physik.uni-mainz.de}

\subjclass[2000]{Primary }

\date{}

\begin{abstract}
There is a fruitful interplay between algebraic geometry on the one side and perturbative quantum field theory on the other side.
I review the main relevant mathematical concepts of periods, Hodge structures and Picard-Fuchs equations and discuss the connection
with Feynman integrals.
\end{abstract}

\maketitle

\section{Periods}

\subsection{Periodic functions}

Let us consider a meromorphic function $f$ of a complex variable $z$.
A period $\omega$ of the function $f$ is a constant such that
\bq
 f\left(z+\omega\right) & = & f\left(z\right)
\eq
for all $z$. Let us exclude the trivial case that $f$ is constant.
The set of all periods of $f$ forms a lattice, which is either trivial, a simple lattice or a double lattice.
The trivial lattice consists of $\omega=0$ only.
A simple lattice is generated by one element. An example for a singly periodic function is given by
\bq
 \exp\left(z\right).
\eq
In this case the simple lattice is generated by $\omega = 2 \pi i$.
A double lattice is generated by two periods $\omega_1$ and $\omega_2$ with
$\mathrm{Im}(\omega_2/\omega_1) \neq 0$.
It is common practice to order these two periods such that $\mathrm{Im}(\omega_2/\omega_1) > 0$.
An example for a doubly periodic function is given by Weierstrass's $\wp$-function.
Let $\Lambda$ be the lattice generated by $\omega_1$ and $\omega_2$
\bq
 \Lambda & = & \left\{ n_1 \omega_1 + n_2 \omega_2 | n_1, n_2 \in {\mathbb Z} \right\}.
\eq
Then
\bq
 \wp\left(z\right)
 & = & 
 \frac{1}{z^2}
 + \sum\limits_{\omega \in \Lambda \backslash \{0\}} \left( \frac{1}{\left(z+\omega\right)^2} - \frac{1}{\omega^2} \right).
\eq
$\wp(z)$ is periodic with periods $\omega_1$ and $\omega_2$.
Of particular interest are also the corresponding inverse functions.
These are in general multivalued functions.
In the case of the exponential function $x=\exp(z)$, the inverse function is given by
\bq
 z & = & \ln\left(x\right).
\eq
The inverse function to Weierstrass's elliptic function $x=\wp(z)$ is an elliptic integral given by
\bq
 z & = &
\int\limits_x^\infty \frac{dt}{\sqrt{4t^3-g_2t-g_3}}
\eq
with
\bq
 g_2 = 60 \sum\limits_{\omega \in \Lambda \backslash \{0\}} \frac{1}{\omega^4},
 & &
 g_3 = 140 \sum\limits_{\omega \in \Lambda \backslash \{0\}} \frac{1}{\omega^6}.
\eq
In both examples the periods can be expressed as integrals involving only algebraic functions.
For the first example of the exponential function we have
\bq
\label{period_integral_1}
 2 \pi i
 & = &
 4 i \int\limits_0^1 \frac{dt}{\sqrt{1-t^2}}.
\eq
For the second example of Weierstrass's $\wp$-function let us assume that $g_2$ and $g_3$ are two given algebraic numbers.
Then
\bq
\label{period_integral_2}
 \omega_1 = 2 \int\limits_{t_1}^{t_2} \frac{dt}{\sqrt{4t^3-g_2t-g_3}},
 & &
 \omega_2 = 2 \int\limits_{t_3}^{t_2} \frac{dt}{\sqrt{4t^3-g_2t-g_3}},
\eq
where $t_1$, $t_2$ and $t_3$ are the roots of the cubic equation $4t^3-g_2t-g_3=0$.

\subsection{Numerical periods}

In the following we will denote by $\mathbb{Q}$ the set of rational numbers and by $\bar{\mathbb Q}$ the set of algebraic numbers.
An algebraic number is a solution of a polynomial equation with rational coefficients.
$\bar{\mathbb Q}$ is a countable subset of ${\mathbb C}$.
Numbers which are not algebraic are called transcendental.
The sets $\mathbb{R}$, $\mathbb{C}$ and the set of transcendental numbers are uncountable.

The representation of the periods of $\exp(z)$ and $\wp(z)$ in the form of eq.~(\ref{period_integral_1}) and eq.~(\ref{period_integral_2})
is the motivation for the following generalisation, due to Kontsevich and Zagier \cite{Kontsevich:2001}:

A numerical period is a complex number whose real and imaginary parts are values
of absolutely convergent integrals of rational functions with rational coefficients,
over domains in $\mathbb{R}^n$ given by polynomial inequalities with rational coefficients.
Domains defined by polynomial inequalities with rational coefficients
are called semi-algebraic sets.

We denote the set of numerical periods by $\mathbb{P}$. 
We may replace in the above definition every occurrence of ``rational function'' with ``algebraic function''
and every occurrence of ``rational number'' with ``algebraic number'' without changing the set of numbers $\mathbb{P}$.
Then it is clear, that the integrals in eq.~(\ref{period_integral_1}) and eq.~(\ref{period_integral_2}) are numerical periods
in the sense of the above definition.

The algebraic numbers are contained in the set of periods, $\bar{\mathbb{Q}} \subset \mathbb{P}$.
The numerical periods $\mathbb{P}$ are a countable set of numbers, lying between $\bar{\mathbb{Q}}$ and $\mathbb{C}$.
The set of numerical periods $\mathbb{P}$ is a $\bar{\mathbb{Q}}$-algebra.
In particular the sum and the product of two numerical periods are again numerical periods.
The transcendental number $\pi$ is contained in $\mathbb{P}$, as can be seen from eq.~(\ref{period_integral_1}).
However it is not known, whether $1/\pi$ belongs to $\mathbb{P}$ or not.
As multiplication with (positive or negative) powers of $(2\pi i)$ is a common operation in number theory, one often considers the algebra
$\mathbb{P}[1/(2 \pi i)]$ with $1/(2 \pi i)$ adjoined.
On the other hand, it is conjectured that the basis of the natural logarithm $e$
and Euler's constant $\gamma_E$ are not periods.

Let us now turn to dimensionally regulated Feynman integrals.
We denote the dimension of space-time by $D$.
We consider a Feynman graph $G$ with $m$ external edges, $n$ internal edges and $l$ loops.
For each internal edge $e_j$ we choose an orientation and associate to the edge 
a momentum $q_j$, a mass $m_j$ and a positive integer number $\nu_j$, giving the power to which the propagator occurs.
At each vertex we impose momentum conservation: 
The sum of all momenta flowing into the vertex equals the sum of all momenta flowing out of the vertex.
Then we can express the momenta flowing through the internal lines as a linear combination of the independent loop momenta
$k_1$, ..., $k_l$ and the external momenta $p_1$, ..., $p_m$ as 
\bq
 q_i & = & \sum\limits_{j=1}^l \rho_{ij} k_j + \sum\limits_{j=1}^m \sigma_{ij} p_j,
 \;\;\;\;\;\; 
 \rho_{ij}, \sigma_{ij} \in \{-1,0,1\}.
\eq
We define the Feynman integral by
\bq
\label{def_feynman_integral_prop}
I_G  & = &
 \frac{\prod\limits_{j=1}^{n}\Gamma(\nu_j)}{\Gamma(\nu-lD/2)}
 \left( \mu^2 \right)^{\nu-l D/2}
 \int \prod\limits_{r=1}^{l} \frac{d^Dk_r}{i\pi^{\frac{D}{2}}}\;
 \prod\limits_{j=1}^{n} \frac{1}{(-q_j^2+m_j^2)^{\nu_j}},
\eq
with $\nu=\nu_1+...+\nu_n$.
$\mu$ is an arbitrary scale introduced to make the integral dimensionless.
After Feynman parametrisation we obtain
\bq
\label{feynman_integral}
I_G  & = &
 \int\limits_{\Delta}  \omega_0
 \left( \prod\limits_{j=1}^n x_j^{\nu_j-1} \right)
 \frac{{\mathcal U}^{\nu-(l+1) D/2}}{{\mathcal F}^{\nu-l D/2}}.
\eq
The prefactors in the defintion of the Feynman integral in eq.~(\ref{def_feynman_integral_prop}) are chosen such that
after Feynman parametrisation we obtain an expression without prefactors, as can be seen from eq.~(\ref{feynman_integral}).
In eq.~(\ref{feynman_integral}) the integration is over
\bq
 \Delta & = & \left\{ \left[ x_1 : x_2 : ... : x_n \right] \in {\mathbb P}^{n-1} | x_i \ge 0, 1 \le i \le n \right\}.
\eq
Here, ${\mathbb P}^{n-1}$ denotes the real projective space with $n-1$ dimensions.
$\omega_0$ is a differential $(n-1)$-form given by
\bq
 \omega_0 & = & \sum\limits_{j=1}^n (-1)^{j-1}
  \; x_j \; dx_1 \wedge ... \wedge \widehat{dx_j} \wedge ... \wedge dx_n,
\eq
where the hat indicates that the corresponding term is omitted.
The functions ${\mathcal U}$ and ${\mathcal F}$ are obtained from first writing
\bq
 \sum\limits_{j=1}^{n} x_{j} (-q_j^2+m_j^2)
 & = & 
 - \sum\limits_{r=1}^{l} \sum\limits_{s=1}^{l} k_r M_{rs} k_s + \sum\limits_{r=1}^{l} 2 k_r \cdot Q_r - J,
\eq
where $M$ is a $l \times l$ matrix with scalar entries and $Q$ is a $l$-vector
with $D$-vectors as entries.
We then have
\bq
\label{definition_U_and_F}
 {\mathcal U} = \mbox{det}(M),
 & &
 {\mathcal F} = \mbox{det}(M) \left( - J + Q M^{-1} Q \right)/\mu^2.
\eq 
${\mathcal U}$ and ${\mathcal F}$ are the first and second graph polynomial of the Feynman graph $G$.
Both polynomials are homogeneous in the Feynman parameters, ${\mathcal U}$ is of degree $l$, ${\mathcal F}$ is of degree $l+1$.
The polynomial ${\mathcal U}$ is linear in each Feynman parameter.
If all internal masses are zero, then also ${\mathcal F}$ is linear in each Feynman parameter.
In the polynomial ${\mathcal F}$ Lorentz invariant quantities like
\bq
 s & = & 
 \left( \sum\limits_j p_j \right)^2
\eq
appear, where the sum runs over a subset of the external momenta. We denote the set of all Lorentz invariants appearing in ${\mathcal F}$
by $S_G$.
There is an alternative definition of the two graph polynomials in terms of spanning forests of the graph $G$ \cite{Bogner:2010kv}.

We would like to discuss one special case of eq.~(\ref{feynman_integral}).
Suppose that
(i) the graph has no external lines or all invariants from $S_G$ are zero,
(ii) all internal masses $m_j$ are equal to $\mu$ and
(iii) all propagators occur with power $1$, i.e. $\nu_j=1$ for all $j$.
Then the Feynman parameter integral reduces to
\bq
\label{localigusazetafct}
I_G  & = &
 \int\limits_{\Delta}  \omega_0 \;
 \left( \sum\limits_{j=1}^n x_j \right)^{l D/2 - n}
 {\mathcal U}^{- D/2}.
\eq
This integral is a Igusa local zeta function (when viewed as a function of $D/2$)
and has been studied by Belkale and Brosnan in \cite{Belkale:2003}.

The Feynman integral defined in eq.~(\ref{feynman_integral}) has an expansion as a Laurent series 
in the parameter $\eps=(4-D)/2$ of dimensional regularisation:
\bq
\label{epsilon_expansion}
 I_G & = &
 \sum\limits_{j=-2l}^\infty c_j \eps^j.
\eq
The Laurent series of an $l$-loop integral can have poles in $\eps$ up to the order $(2l)$. The poles in $\eps$ correspond to
ultraviolet or infrared divergences.
The coefficients $c_j$ are functions of the Lorentz invariants $s \in S_G$, the masses $m_i$ and (in a trivial way) of the arbitrary scale $\mu$.
Suppose for the moment that (i) all kinematical invariants $s \in S_G$ are negative or zero, 
(ii) all masses $m_i$ and $\mu$ are positive or zero ($\mu\neq0$) and
(iii) all ratios of invariants and masses are rational,
then it can be shown that all coefficients $c_j$
of the Laurent expansion are numerical periods.
For the special case of Igusa local zeta functions in eq.~(\ref{localigusazetafct}) this has been proven
by Belkale and Brosnan \cite{Belkale:2003},
the more general case of eq.~(\ref{feynman_integral}) in \cite{Bogner:2007mn}.
The non-trivial part of this statement comes from the fact that in four space-time dimensions Feynman integrals 
can diverge. A value of $D \neq 4$ acts as a regulator.
In general the expansion in $\eps$ does not commute with the integration.
In order to be able to write the coefficients $c_j$ as manifestly absolute convergent integrals one uses a resolution of the singularities
of the two graph polynomials \cite{Hironaka:1964,Spivakovsky:1983}.
In the physics community a constructive algorithm to do this is known as sector decomposition \cite{Binoth:2000ps,Bogner:2007cr}.

\subsection{Abstract periods}
\label{subsection_abstract_periods}

There is a more formal definition of periods as follows \cite{Kontsevich:2001}:
Let $X$ be a smooth algebraic variety of dimension $n$ defined over $\mathbb{Q}$
and $D \subset X$ a divisor with normal crossings.
(A normal crossing divisor is a subvariety of dimension $n-1$, which looks locally like a union of coordinate hyperplanes.)
Further let $\omega$ be an algebraic differential form on $X$ of degree $n$
and $\Delta$ a singular $n$-chain on the complex manifold $X(\mathbb{C})$ with boundary on the divisor $D(\mathbb{C})$.
We thus have a quadruple $(X,D,\omega,\Delta)$. To each quadruple we can associate a complex number $P(X,D,\omega,\Delta)$ 
called the period of the quadruple and 
given by the integral
\bq
P(X,D,\omega,\Delta) & = & \int\limits_\Delta \omega.
\eq
It is clear that the period of the quadruple is an element of $\mathbb{P}$, and that to any element $p \in \mathbb{P}$ one can find a quadruple,
such that $P(X,D,\omega,\Delta)=p$.
The period map is therefore surjective. The interesting question is whether the period map is also injective.
As it stands above, the period map is certainly not injective for trivial reasons.
For example, a simple change of variables can lead to a different quadruple, but does not change the period.
One therefore considers equivalence classes of quadruples modulo relations induced by linearity in $\omega$ and $\Delta$,
changes of variables, Stokes' formula and Fubini's formula.
The vector space over $\mathbb{Q}$ of these equivalence classes is called the space of effective periods and denoted by ${\mathcal P}$.
${\mathcal P}$ is an algebra.
It is conjectured that the period map from ${\mathcal P}$ to $\mathbb{P}$ is injective and therefore an isomorphism \cite{Grothendieck:1966,Kontsevich:2001,Ayoub:2011}.
This would imply that all relations between numerical periods are due to linearity, change of variables, Stokes and Fubini.

In order to make the definition more concrete, we consider an an example the quadruple given by
$X = {\mathbb A}_{\mathbb Q}^1 \backslash \{0\}$, $D=\emptyset$, $\omega = dz/z$ and $\Delta$ the path along the unit circle in the counter-clockwise
direction.
The corresponding period is $(2\pi i)$.

As in the case of numerical periods it is not known whether there is a quadruple in ${\mathcal P}$, whose period is $1/(2 \pi i)$.
One therefore adjoins to ${\mathcal P}$ formally the inverse of the element whose period is $(2\pi i)$ and writes
${\mathcal P}[1/(2 \pi i)]$ for the so obtained algebra.
Elements of ${\mathcal P}[1/(2 \pi i)]$ are called abstract periods.

\section{Hodge structures}

Hodge structures have their origin in the study of compact K\"ahler manifolds.
Let $M$ be a complex manifold with complex structure $J$.
A Riemannian metric $g$ on $M$ is called Hermitian, if it is compatible with the complex structure $J$, in other words
\bq
 g\left( J X, J Y \right) & = & g\left( X, Y \right).
\eq
$X$ and $Y$ denote vector fields on $M$.
For a Hermitian manifold one defines an associated differential two-form by
\bq
 K\left(X,Y\right) & = & g\left( J X, Y \right).
\eq
$K$ is called the K\"ahler form. (In the literature also the letter $\omega$ is frequently used to denote the K\"ahler
form.)
A Hermitian manifold is called a K\"ahler manifold, if the two-form $K$ is closed:
\bq
 d K & = & 0.
\eq
An example of compact K\"ahler manifolds is provided by compact Riemann surfaces.
Riemann surfaces are complex manifolds of complex dimension one and the K\"ahler form of any Hermitian metric is necessarily closed.
A second example is given by the complex projective space $\mathbb{P}^n(\mathbb{C})$.
As a third example we mention complex submanifolds of K\"ahler manifolds. These submanifolds are again K\"ahler.

On a compact K\"ahler manifold we have the following decomposition of the cohomology groups
\bq
 H^k\left(X\right) \otimes {\mathbb C}& = & \bigoplus\limits_{p+q=k} H^{p,q}(X),
 \;\;\;\;\;\;
 \overline{H^{p,q}(X)} = H^{q,p}(X).
\eq
For a fixed $k$ this provides an example of a pure Hodge structure of weight $k$.
A pure Hodge structure of weight $k$ on a $\mathbb Z$-module $V_{\mathbb Z}$ of finite rank
is a direct sum decomposition
\bq
 V_{\mathbb C} & = & V_{\mathbb Z} \otimes_{\mathbb Z} {\mathbb C}
 =
 \bigoplus\limits_{p+q=n} V^{p,q}
 \;\;\;
 \mbox{with}
 \;\;\;
 \overline{V^{p,q}} = V^{q,p}.
\eq
If one replaces $\mathbb Z$ by $\mathbb Q$ or $\mathbb R$, one speaks about a rational or real Hodge structure, respectively.
The numbers
\bq
 h^{p,q}(V) & = & \mbox{dim}\;V^{p,q}
\eq
are called the Hodge numbers.
There is a second definition of a pure Hodge structure, which is more adapted for generalisations.
The second definition is based on a Hodge filtration: Let $F^\bullet V_{\mathbb C}$ be 
a finite decreasing filtration:
\bq
 V_{\mathbb C} \supseteq ... \supseteq F^{p-1} V_{\mathbb C} \supseteq F^{p} V_{\mathbb C} \supseteq F^{p+1} V_{\mathbb C} \supseteq ... \supseteq (0)
\eq
such that
\bq
 V_{\mathbb C} & = &
 F^p V_{\mathbb C} \oplus \overline{F^{k-p+1} V_{\mathbb C}}.
\eq
Then $V$ carries a pure Hodge structure of weight $k$.
These two definitions are equivalent: Given the Hodge decomposition, we can define the
corresponding Hodge filtration by
\bq
 F^{p} V_{\mathbb C} & = & \bigoplus\limits_{j \ge p} V^{j,k-j}.
\eq
Conversely, given a Hodge filtration we obtain the Hodge decomposition by
\bq
 V^{p,q} & = & F^p V_{\mathbb C} \cap \overline{F^q V_{\mathbb C}}.
\eq
Pure Hodge structures are relevant for smooth projective algebraic varieties, these are necessarily compact.
If one gives up the requirement of smoothness or compactness one is lead to a 
generalisation called mixed Hodge structure \cite{Deligne:1970,Deligne:1971,Deligne:1974}.
A mixed Hodge structure is given by
a $\mathbb Z$-module $V_{\mathbb Z}$ of finite rank,
a finite increasing filtration on 
$V_{\mathbb Q} = V_{\mathbb Z} \otimes_{\mathbb Z} {\mathbb Q}$, called the weight filtration:
\bq
 (0) \subseteq ... \subseteq W_{k-1} V_{\mathbb Q} \subseteq W_{k} V_{\mathbb Q} \subseteq W_{k+1} V_{\mathbb Q} \subseteq ... \subseteq V_{\mathbb Q},
\eq
and a finite decreasing filtration on
$V_{\mathbb C} = V_{\mathbb Z} \otimes_{\mathbb Z} {\mathbb C}$, called the Hodge filtration:
\bq
 V_{\mathbb C} \supseteq ... \supseteq F^{p-1} V_{\mathbb C} \supseteq F^{p} V_{\mathbb C} \supseteq F^{p+1} V_{\mathbb C} \supseteq ... \supseteq (0),
\eq
such that $F^\bullet$ induces a pure Hodge structure of weight $k$ on
\bq
 \mbox{Gr}_k^W V_{\mathbb Q} & = & {W_k V_{\mathbb Q}} / {W_{k-1} V_{\mathbb Q}}.
\eq
In addition, we can consider a family of Hodge structures, parametrised by a manifold.
If in addition a few technical conditions are met, this leads to a variation of a Hodge structure.
The precise definition of a variation of a Hodge structure can be found in the literature \cite{Griffiths:1968i,Griffiths:1968ii}. 
Suppose first that the variation with the parameters is smooth.
Then the Hodge numbers $h^{p,q}$, being integers, have to remain constant.
If the variation with the parameters has singularities, one can study what happens as one approaches the singularity.
This gives a limiting mixed Hodge structure.
For an exact definition the reader is again referred to the literature \cite{Schmid:1973,Steenbrink:1976}.

In order to discuss the connection of Hodge structures with periods we focus on a 
smooth 
algebraic variety over the rational numbers ${\mathbb Q}$.
To this variety we can on the one hand associate the de Rham cohomology $H_{\mathrm{DR}}^k(X)$, as well as
the Betti cohomology $H_{\mathrm{B}}^k(X)$. 
Let us denote by $\omega_j$ a basis for $H_{\mathrm{DR}}^k(X)$ and by $\gamma_i$ a basis for the Betti homology $H^{\mathrm{B}}_k(X)$.
A basis for $H_{\mathrm{B}}^k(X)$ is then given by the duals $\gamma_j^\ast$ and satisfies $\gamma_i \gamma_j^\ast = \delta_{ij}$. 
There is an isomorphism between the de Rham and Betti cohomology,
\bq
 H_{\mathrm{B}}^k(X) \otimes \mathbb{C} \rightarrow H_{\mathrm{DR}}^k(X) \otimes \mathbb{C}
\eq
given by
\bq
 \omega_i = \sum\limits_j p_{ij} \gamma_j^\ast,
 & &
 p_{ij} = \int\limits_{\gamma_j} \omega_i.
\eq
The coefficients $p_{ij}$ are called the periods of $X$.
The abstract periods are then the equivalence classes of the quadruples $(X,\emptyset,\omega_i,\gamma_j)$.

\begin{figure}
\includegraphics[bb= 270 630 360 700, scale=.65]{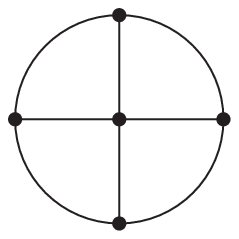}
\includegraphics[bb= 225 630 405 720, scale=.65]{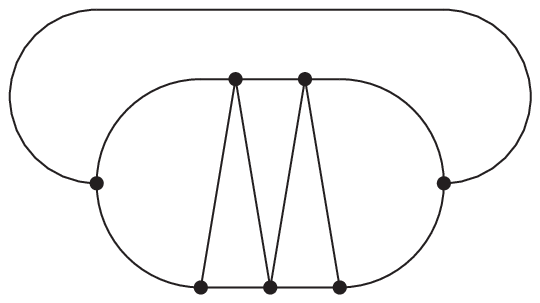}
\includegraphics[bb= 250 630 360 700, scale=.65]{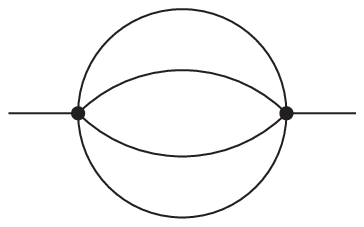}
\caption{The wheel with four spokes (left), the $6$-loop zigzag graph (middle) and the $3$-loop banana graph (right).}
\label{fig1}
\end{figure}

Let us now come back to the Feynman integral in eq.~(\ref{feynman_integral}) and focus on finite Feynman integrals, which do not need
to be regulated.
We consider two special cases:
The first case is given by $D=4$, $\nu_j=1$ and $n=2l$. Then
\bq
\label{example_U}
I_G  & = &
 \int\limits_{\Delta}  \frac{\omega_0}{{\mathcal U}^{2}}.
\eq
Examples for this case are the wheel with $l$ spokes and the family of zigzag graphs \cite{Bloch:2006,Doryn:2010zz,Brown:2012ia}. 
Examples of corresponding Feynman graphs are shown in fig.~(\ref{fig1}) on the left and in the middle, respectively.
In the second case we consider $D=2$, $\nu_j=1$ and $n=l+1$. Then
\bq
\label{example_F}
I_G  & = &
 \int\limits_{\Delta}  
 \frac{\omega_0}{{\mathcal F}}.
\eq
Examples for this case are the family of banana graphs \cite{Aluffi:2008sy,MullerStach:2011ru}.
An example of a corresponding Feynman graph is shown in fig.~(\ref{fig1}) on the right.

In both cases the integrand is a rational differential form.
We would like to think of these integrals as periods of mixed Hodge structures.
There are two complications to this. First of all, the integrand becomes singular when the polynomials in the denominator vanish.
Secondly, the region of integration $\Delta$ has a boundary $\partial \Delta$.
It is therefore clear, that there are two geometric objects of interest
in eq.~(\ref{example_U}) or eq.~(\ref{example_F}):
On the one hand the domain of integration $\Delta$ and on the other hand the algebraic variety $X$ defined by the zero set of 
the polynomial in the denominator.
In the first example, $X$ is defined by ${\mathcal U}^2=0$, in the second example $X$ is defined by ${\mathcal F}=0$.
The two objects $X$ and $\Delta$ may intersect.
Since ${\mathcal U}$ is a sum of monomials with all coefficients equal to $1$, intersections of $X$ and $\Delta$ may happen in the case of
the first example only on the boundary $\partial \Delta$.
The same holds true in the second example if we restrict ourselves for the kinematics to the Euclidean region, already encountered
after eq.~(\ref{epsilon_expansion}):
If (i) all kinematical invariants $s \in S_G$ are negative or zero and (ii) all masses $m_i$ and $\mu$ are positive or zero ($\mu\neq0$),
then the zero set of ${\mathcal F}$ intersects $\Delta$ only on the boundary.
If $X \cap \partial \Delta$ is non-empty, we blow-up $\mathbb{P}^{n-1}$ in this region.
Let us denote the blow-up by $P$.
We further denote the strict transform of $X$ by $Y$ and the total transform of the set 
$\{ x_1 x_2 ... x_n = 0 \}$ by $B$.
With these notations we can now consider the mixed Hodge structure (or the motive) given by the relative cohomology group \cite{Bloch:2006,Bloch:2008jk,Connes:2005zp,Aluffi:2009aa,Brown:2009a,Schnetz:2008mp}
\bq
 H^{n-1}\left(P \backslash Y, B \backslash B \cap Y \right).
\eq
The Feynman integral is then a period of this cohomology class.
Let us denote by
\bq
 S & = & S_G \cup \{m_1^2,...,m_n^2\}
\eq
the set of all Lorentz invariants $s$ and masses squared appearing in ${\mathcal F}$.
The polynomial ${\mathcal F}$ depends on these variables and we actually have in the example of eq.~(\ref{example_F})
a variation of a mixed Hodge structure.

\section{Picard-Fuchs equations}

Do the formal considerations of the previous section have practical applications~? Yes, they do.
We will give an example.
Periods of variations of mixed Hodge structures are expected to satisfy differential equations of Picard-Fuchs type.
In the case of example~(\ref{example_F}) the Feynman integral is a period of a variation of a mixed Hodge structure and
should satisfy for any given choice $t \in S$ an ordinary linear differential equation of Picard-Fuchs type.
Such a differential equation is a useful tool for computing the Feynman integral, once the boundary values and the inhomogeneous terms
in the differential equation are known.
Below we will sketch an algorithm for finding for a given variable $t \in S$ an ordinary linear differential equation of Picard-Fuchs type
of minimal order \cite{MullerStach:2012mp}.
The algorithm has the additional benefit that we can work with dimensionally regulated integrals as in eq.~(\ref{feynman_integral})
and are no longer forced to restrict ourselves to rational integrands as in eq.~(\ref{example_U}) or in eq.~(\ref{example_F}).

Starting from eq.~(\ref{feynman_integral}) we pick one $t \in S$ and set
\bq
 \omega_t =  
 f \omega_0,
 & &
 f =
 \left( \prod\limits_{j=1}^n x_j^{\nu_j-1} \right)
 \frac{{\mathcal U}^{\nu-(l+1) D/2}}{{\mathcal F}^{\nu-l D/2}}.
\eq
We look for a differential equation of the form
\bq
\label{basic_picard}
 L^{(r)} \omega_t & = & d \beta,
\eq
where
\bq
\label{picard_fuchs_operator}
 L^{(r)} & = & \sum\limits_{j=0}^r p_j \left( \mu^2 \frac{d}{dt} \right)^j
\eq
is a Picard-Fuchs operator of order $r$.
The coefficients $p_j$ may depend on the kinematical invariants from the set $S$, the scale $\mu$, the space-time
dimension $D$ and the exponents $\nu_i$, but not on the Feynman parameters $x_i$. We normalise the Picard-Fuchs operator such that $p_r=1$.
We recall that a differential operator of the form as in eq.~(\ref{picard_fuchs_operator}) is said to be of Fuchsian type, if all coefficients $p_j$ are
meromorphic functions of $t$ and if $p_j$ has at most poles of order $(r-j)$.
$\beta$ is a $(n-2)$-form, depending on the Feynman parameters $x_i$. The differential $d$ is with respect to the Feynman parameters $x_i$.
For $\beta$ we make the following ansatz
\bq
\label{ansatz_beta}
 \beta & = & 
 \frac{f}{{\mathcal F}^{r-1}}
 \sum\limits_{j_1<j_2} \left(-1\right)^{j_1+j_2}
 \left[ -x_{j_1} a_{j_2} + x_{j_2} a_{j_1} \right]
 \\
 & &
 \nonumber
 dx_1 \wedge ... \wedge \widehat{dx_{j_1}} \wedge ... \wedge \widehat{dx_{j_2}} \wedge ... \wedge dx_n,
\eq
where the $a_i$ are homogeneous polynomials of degree $(r l + r - l)$
in the variables $x_i$.
The ansatz is based on the fact, that the singularities of the integrand $\omega_t$ 
are given by powers of the graph polynomials ${\mathcal U}$ and ${\mathcal F}$.
Acting with $L^{(r)}$ on the integrand $\omega_t$ 
will only increase the power of ${\mathcal F}$ in the denominator by $r$, but will not introduce singularities on new algebraic varieties.
Each polynomial $a_i$ contains only a finite number of monomials in the Feynman parameters with a priori unknown coefficients.
We therefore take these coefficients and the variables $p_0$, ..., $p_{r-1}$ as the set of our unknown variables.
Plugging the ansatz~(\ref{ansatz_beta}) in eq.~(\ref{basic_picard}) gives a linear system of equations for the unknown variables.
This system may or may not have a solution.
In order to find the differential equation of minimal order we start at $r=1$ and try to solve the linear system of equations.
If no solution is found, we increase $r$ by one and repeat the exercise, until a solution is found.
This is then the solution of minimal order $r$.

Integration of eq.~(\ref{basic_picard}) yields then
\bq
\label{differential_equation}
 L^{(r)} I_G & = & \int\limits_\Delta d \beta
 =
\int\limits_{\partial \Delta} \beta,
\eq
where we used Stokes.
Eq.~(\ref{differential_equation}) is the sought-after ordinary linear differential equation for the Feynman integral $I_G$.
The right-hand side is given as a sum of integrals with $(n-1)$ Feynman parameters.
These integrals correspond to graphs with one propagator less and can be considered as simpler.

To give an example we consider the one-loop two-point function with equal internal masses in two space-time dimensions
This integral is given by
\bq
 I_G & = & \int\limits_{\Delta} \frac{x_1 dx_2-x_2 dx_1}{{\mathcal F}},
 \;\;\;\;\;\;
 {\mathcal F} = - \frac{t}{\mu^2} x_1 x_2 + \left( x_1 + x_2 \right)^2 \frac{m^2}{\mu^2}.
\eq
This integral has a first-order Picard-Fuchs operator
\bq
 L^{(1)} & = & \mu^2 \frac{d}{dt} + \mu^2 \frac{t-2m^2}{t(t-4m^2)}
\eq
and a possible solution for $\beta$ is
\bq
 \beta & = & \frac{1}{\mathcal F} \frac{\mu^2}{t(t-4m^2)} \left[ (t-2m^2) x_1 x_2 - 2 m^2 x_2^2 \right].
\eq
The boundary of $\Delta$ is given by the two points $[1:0]$ and $[0:1]$. 
The integration of the inhomogeneous term yields
\bq
\int\limits_{\partial \Delta} \beta & = & - \frac{2\mu^4}{t(t-4m^2)}.
\eq
Putting everything together, we obtain the differential equation
\bq
 \left( t(t-4m^2) \frac{d}{dt} + t-2m^2 \right) I_G & = & - 2\mu^2.
\eq


\providecommand{\bysame}{\leavevmode\hbox to3em{\hrulefill}\thinspace}
\providecommand{\MR}{\relax\ifhmode\unskip\space\fi MR }
\providecommand{\MRhref}[2]{%
  \href{http://www.ams.org/mathscinet-getitem?mr=#1}{#2}
}
\providecommand{\href}[2]{#2}

\end{document}